
\magnification=1200
\baselineskip=19pt
\overfullrule=0pt
\centerline{\bf NEW SETS OF KINK BEARING HAMILTONIANS}

\vskip .8in

\centerline{B.~ Dey }
\centerline{ Department of Physics, University of Poona,
Ganeshkhind, Pune 411 007, India.}

\centerline{ C.~N.~ Kumar \footnote *{
E-mail address: cnkumar@cts.iisc.ernet.in}}

\centerline{Centre for Theoretical Studies, Indian
Institute of Science, Bangalore 560012, India.}

\vskip .5in

\centerline{\bf ABSTRACT }
\vskip .5in
Given a kink bearing Hamiltonian, Isospectral Hamiltonian
approach is used in generating new sets of Hamiltonains which
also admit kink solutions.  We use Sine-Gordon model as a
example and explicitly work out the new sets of potentials and
the solutions.

\vfill
\eject

\vskip .2in
Nonlinear field theory models in (1+1) dimensions, with
Lagrangian density
\break
${\cal L}={1\over 2}
\partial_{\mu}\phi\partial^{\mu}\phi~-V(\phi)$ , for which the
equations of motion admit finite energy, finite width solutions
have been well studied $^1$.  Sine-Gordon model and $\phi^4$
models are two popular models used for modelling some physical
systems.  While the solutions of $\phi^4$ are called
kinks/lumps, the solutions of Sine-Gordon model have the extra
property that they retain their identity after collisions, and
they are called solitons. In order to model a variety of
physical systems, various attempts are made to enlarge the class
of nonlinear field theoretic models. Parametrically modified
Sine-Gordon model was one of such earlier attempts made by
Remoissenet and Peyrard $^2$.  In this model the potential is
$V(\phi,r)$ whose shape can be varied continuosly as a function
of $r, ~~-1~<~r~<~1$ and has the Sine-Gordon (SG) potential for
$r=0$ as a special case. The implicit kink solutions for this
model and their rest masses are calculated. For $r \ne 0$ the
model is not completely integrable. Kink-antikink interactions
are studied for this model and was shown that the structure is
similar to that observed in $K\bar K$ interactions for the
$\phi^4$ model for a range of $r$ $^3$.

In this article we present a prescription to construct a family
of potentials which admit kink solutions. The method is based on
Isospectral Hamiltonian approach which enables us to generate
two sets of potentials from a given potential. In one case we
can give explicit kink solution whereas implicit solutions are
obtained in the other case. A partial result of this approach
using $\phi^4$ model was presented earlier by one of us $^4$.

  Once the field theoretic model admits kink type solutions, the
stability of the kink is ensured by the occurence of the zero
energy ground state of the stability equation when small
oscillations around the kink are considered $^1$. Considering
the stability equation as a one dimensional Schrodinger-like
equation for a particle in a potential $V(x)$ we can construct a
isospectral partner for it.  Then, as we explain below,
following the work of Christ and Lee $^5$ we shall construct the
kink solution and the potential which admits the solution from
the partner stability equation.

Let the Lagrangian density of a single hermitian scalar field
$\phi$ in 1+1 dimensions be given by $${\cal L}={1\over 2}
\partial_{\mu}\phi\partial^{\mu}\phi~-V(\phi)\eqno (1)$$
\noindent In order to have kink solutions the potential $V(\phi)$ is
assumed to have atleast two degenerate absolute minima. The time
independent field equations reads $${{d^2\phi}\over
{d{x^2}}}-{dV\over{d\phi}}=0\eqno(2)$$
\noindent which can also be written as
$${1\over 2}{({{d\phi}\over {d x}})^2}=V(\phi)\eqno(3)$$
\noindent on taking squareroot of equn (3) and integrating
$$\int
{\lbrack\sqrt{2V(\phi)}\rbrack}^{-1}~d\phi~=~x+{x_0}\eqno(4)$$
\noindent This relation gives $\phi$ in terms of $x$ an implicit
solution. For the cases where $\phi$ can be expressed in terms
of $x$ the explicit solutions can be obtained. The stability of
the kink solution is ensured by the occurence of the zero energy
ground state of the corresponding stability equation. Expanding
the $\phi$ around kink solution as $\phi~={\phi_k}~+\psi$ one
gets $$\lbrack -{d^2\over {d x^2}}+V^{\prime\prime}
(\phi_k)\rbrack {\psi_n} ={E_n}{\psi_n}\eqno(5)$$
\noindent It is straight forward to see that
$${E_0}=0,~~ {\rm and}~~ {\psi_0}={d{\phi_k}\over
{dx}}\eqno(6)$$ satisfies the equation (5). Infact the relation
between the $\psi_{0}$ and $\phi_{k}$ is the result of
translational invariance of the Lagrangian, field equation and
the solution $^1$.  The equations (3) and (6) play an important
role in our analysis.

We can consider the stability equation as a 1-dimensional
Schordinger like equation for a particle in 1-dimensional
potential. The knowledge of the ground state energy and the
ground state wavefunction enables to factorise the equation.
Once the factorisation is done as $H={A^+}{A^-}+\epsilon_{0}$
the energy spectrum of the $H$ and its partner
${H_1}={A^-}{A^+}+\epsilon_{0}$ are related by supersymmetry as
$E_{1}^{(n)}=E^{n+1}, (n=0,1,2,3 ...)$ where $E,~ {E_1}$ are the
energy spectrum of $H,~ {H_1}$ respectively $^6$.  The operators
$A^{\pm}$ are given by ${1\over\sqrt
2}{(\pm{\partial\over{\partial x}} +\alpha(x))},$ where
$\alpha(x)={d\over {dx}}\log\psi_{0}$, $\psi_{0}$ is the ground
state wavefunction of $H$. Given a parent Hamiltonian $H$ with
$N$ energy levels, one can construct the 'daughter' Hamiltonian
$H_1$ with $N-1$ levels and the procedure can be iterated. At
the $H_1$ level the question can be asked, following Mielnek
$^7$  whether the factorisation of $H_1$ is unique or not.
Consider $ {H_1}={B^-}{B^+}+\epsilon_{0}$ is another
factorisation where $B^{\pm} ={1\over\sqrt
2}(\pm{\partial\over{\partial x}} +\beta(x))$.  An obivous
particular solution is $\beta(x)=\alpha(x)$. Let the general
solution be $\beta(x)=\alpha(x)+\chi(x)$ This yields
$${\chi^2}(x)+2\chi(x)\alpha(x)-{\chi^\prime}(x)=0$$
\noindent This is a Riccatti equation whose solution is
$$\chi(x)=\psi_{0}^2~(C-\int \psi_{0}^{2}dx)^{-1}$$
\noindent where $\alpha(x)={d\over {dx}}\log\psi_{0}$, and $C$ is a
integration constant chosen such that $\chi$ is non-singular.
Hence we have another factorisation i.e.
${H_1}={A^-}{A^+}+\epsilon_{0}={B^-}{B^+}+\epsilon_{0}$. At this
stage the above equation is little new to offer. However if we
construct ${B^+}{B^-}$ it is no longer $H$ but a new Hamiltonian
$H^{N}$ $$H^{N}=H_{1}+{\partial\over{\partial x}}\beta(x)$$
\noindent We call $H^N$ an isospectral partner of $H$ in the
sense that it has same energy spectrum as that of $H$. The
ground state wave of $H^N$ is given by
$${\psi_{0}^N}=\psi_{0}{{(C-\int{\psi_{0}^2}dx)}^{-1}}\eqno(7)$$
\noindent In a sense nonuniqueness of factorising $H$ has led us
to construct one more parent Hamiltonian $H^N$ in addition to
the original Hamiltonian $^4$.

Now that the partner stability equation has been constructed,
following the ref. 5 (using equns. 3 and 6) the kink solution
and the corresponding field theory admitting the solution can be
obtained, but a slight subtlety enables us to generate two sets
of Hamiltonians as we explain here. In what follows we shall
keep the discussion general as well work out the Sine-Gordon
model as a specific example.  Using the equn. (6) in equn.(7)
and integration gives $${\phi^{N}_{k}}=\int
{\psi_{0}{(C-\int{\psi_{0}^2}dx)}^{-1}}dx\eqno(8)$$ again to
remind $C$ is an integration constant earlier chosen such that
$\chi$ is nonsingular. $\phi^{N}_{k}$ is the new kink solution
and the potential is given by the equation (3).

For the specific case of SG potential with $m=1$,
$V(\phi)=(1-\cos\phi)$ and the soliton solution is
$4\tan^{-1}\exp~{x}$.  The zero energy translational mode reads
$\psi_{0}=sech x$ therefore $$\psi_{0}^{N}={sech x\over{(C-tanh
x)}}~~,~~ \vert C \vert > ~1$$
\noindent integrating this equation gives
$$\phi_{k}^{N}(x)={1\over{\sqrt {C^2
-1}}}\tan^{-1}(sinh(x-x_{0})), ~~~~x_{0}=\tanh^{-1}({1\over
C})\eqno(9)$$
\noindent The kink solution asymptotically $(x\rightarrow \pm\infty )$
takes  the values $\pm {\pi\over{\sqrt({C^2}-1)}}$.
\noindent On using the equn.(3) the potential takes the form
$$V(\phi)={1\over{2({C^2}-1)}}{\cos^2}({\sqrt{C^2-1}}~\phi)\eqno(10)$$

It has to be mentioned here that, for the Sine-Gordon example
the new potential and the kink solution are just rescaled and
translated versions of the original potential and kink solution
respectively.  In the case of $\phi^4$ model this procedure does
lead to different result.  But the algebraic complexity of
expressing $x$ in terms of $\phi$ and thus $V(\phi)~({1\over
2}(d\phi/dx)^2)$ as a function of $\phi$ hindered the further
study of the potential $^4$.

In the second case, the starting point would be to use the
equns. (3,6,7) together instead of integrating equn. (7) in
conjunction with equn.(6).  We have $${1\over 2}{({d\phi\over
{dx}})^2}=V(\phi)\eqno(3^\prime)$$ $$\psi_{0}={d\phi\over
{dx}}\eqno(6^\prime)$$
$$\psi_{0}^N={\psi_{0}{{(C-\int{\psi_{0}^2}dx)}^{-1}}}\eqno(7^\prime)$$
\noindent Substituting$(6^\prime)$ in $(7^\prime)$ and squaring the
resulting equation it is straight forward to see that
$$V(\phi^{N})={V(\phi)\over(C-\int \sqrt{2 V(\phi)}~d\phi)^2
}\eqno(11)$$ In the case of SG model $V(\phi)=(1-\cos\phi)$ and
the $V^{N}(\phi)$ reads $$V^{N}(\phi)={1\over
16}{(1-\cos\phi)\over{({C\over 4}+\cos\phi/2)^{2}}} ~~,~~~~\vert
C \vert > 4\eqno(12)$$
\noindent The kink solution in this case is an implicit one and
has the form $$2~\tan^{{C\over 4}-1}({\phi\over
4})~{\sin^{2}({\phi\over 4})}=
\exp{(x+x_{0})\over 4}\eqno(13)$$
\noindent The asymptotic behaviour is as follows. As $x\rightarrow\infty,~~
\phi=2\pi~~{\rm as}~~x\rightarrow -\infty,~~\phi=0$.
The potential in this case has $4\pi$ periodicity, therefore two
types of kink solutions exist. One was described above and the
second solution $\phi_{2}(x)$ takes the values $(2\pi,4\pi)~~
{\rm as}~~ x\rightarrow
\pm\infty$  and the solution is given by $\phi_{2}(x)=4\pi-\phi_{1}(-x)$.
The representative profiles of the  potential and the kink
solution are plotted for three values of $C$ (see Figs.1 and 2).
It is interesting to note that while minima of the potential are
pegged at the same point the maxima are dependent on the
parameter $C$.

Two remarks are in order. (1). While in the first case the kink
solution is obtained first using equn.(6) and the potential is
derived using equn.(3), whereas in the second case the potential
is obtained first and then the solution is derived using the
standard integration, resulting two different sets of kink
bearing Hamiltonians. (2). The potential for the SG model in the
second case is different when compared to the case considered in
ref.2 $V(\phi)$ in their case is given by
$$V(\phi)={(1-r)^{2}}{(1-\cos\phi)\over{(1+{r^2}+2r\cos\phi)}}~~,~~\vert
r
\vert< 1\eqno(14)$$
SG case can be obtained in their case for the value $r=0$.

In conclusion, using the Isospectral Hamiltonian approach, we
constructed two new sets of Hamiltonians from a given
Hamiltonian which admits kink solutions. SG model has been
worked out as a specific example. We feel the second way of
generating potentials is useful in the light of Peyrard and
Campbell work on kink-antikink interactions, earlier referred to
in this article. A related problem in these cases is to examine
if the present kinks are of soliton or quasi-soliton type.

\bigskip
{\bf Acknowledgement}

One of us (C.N.K.) thanks B.~Dutta of JNCASR for his help.

\bigskip
\noindent{\bf REFERENCES}

\noindent
1. R. Rajaraman  (1982) Solitons and Instantons (Amsterdam:
North Holland)

\noindent
2. M. Remoissenet  and M. Peyrard  (1981) J.Phys. C.{\bf 14}
L481.

\noindent
3. M. Peyrard  and D.K. Campbell  (1983) Physica {\bf 9D} 33.

\noindent
4. C.N. Kumar  (1987) J.Phys.A. Math.Gen.{\bf 20} 5397.

\noindent
5. N.H. Christ  and T.D. Lee  1975 Phys. Rev.  {\bf D12} 1606.

\noindent
6. C.V. Sukumar (1985) J.Phys.A.Math.Gen {\bf 18} L57, 2917,
2937.

\noindent
7. B. Mielnik  (1984) J.Math.Phys. {\bf 25} 3387.
\bigskip

\centerline{\bf FIGURE CAPTIONS}
\smallskip
Fig.~1. The potential $V^N (\phi)$ (equn. 12) for (a) C=5 (b)
C=7 and (c) C=9

Fig.~2. The kink profile (equn. 13) for (a) C=5 (b) C=7 and (c)
C=9.
\vfil
\eject
\end